\documentclass[onecolumn, 11pt]{IEEEtran}
\usepackage{amsmath}
\usepackage{amssymb}
\usepackage{amsfonts}
\usepackage{graphicx}
\usepackage{epsfig}
\usepackage{subfigure}
\usepackage{psfrag}
\usepackage{cite}
\usepackage{latexsym}
\usepackage{url}
\usepackage{color}

\begin{document}
\title{Cost-Aware Green Cellular Networks with Energy and Communication Cooperation}
\author{Jie Xu, Lingjie Duan, and Rui Zhang
\thanks{J. Xu and R. Zhang are with the Department of Electrical and Computer Engineering, National
University of Singapore (e-mail: elexjie@nus.edu.sg, elezhang@nus.edu.sg).}
\thanks{L. Duan is with the Engineering Systems and Design Pillar, Singapore University of Technology and Design (e-mail: lingjie\_duan@sutd.edu.sg).}
}

\setlength{\textwidth}{7.1in} \setlength{\textheight}{9.7in}
\setlength{\topmargin}{-0.8in} \setlength{\oddsidemargin}{-0.30in}

\maketitle

\begin{abstract}
Energy cost of cellular networks is ever-increasing to match the surge of wireless data traffic, and the saving of this cost is important to reduce the operational expenditure (OPEX) of wireless operators in future. The recent advancements of renewable energy integration and two-way energy flow in smart grid provide potential new solutions to save the cost. However, they also impose challenges, especially on  how to use the stochastically and spatially distributed renewable energy harvested at cellular base stations (BSs) to reliably supply time- and space-varying wireless traffic over cellular networks. To overcome these challenges, in this article we present three approaches, namely, {\emph{energy cooperation, communication cooperation, and joint energy and communication cooperation}}, in which different BSs bidirectionally trade or share energy via the aggregator in smart grid, and/or share wireless resources and shift loads with each other to reduce the total energy cost.
\end{abstract}

\setlength{\baselineskip}{1.3\baselineskip}
\newtheorem{definition}{\underline{Definition}}[section]
\newtheorem{fact}{Fact}
\newtheorem{assumption}{Assumption}
\newtheorem{theorem}{\underline{Theorem}}[section]
\newtheorem{lemma}{\underline{Lemma}}[section]
\newtheorem{corollary}{\underline{Corollary}}[section]
\newtheorem{proposition}{\underline{Proposition}}[section]
\newtheorem{example}{\underline{Example}}[section]
\newtheorem{remark}{\underline{Remark}}[section]
\newtheorem{algorithm}{\underline{Algorithm}}[section]
\newcommand{\mv}[1]{\mbox{\boldmath{$ #1 $}}}

\section{Introduction}\label{sec:1}

To meet the dramatically increasing mobile data traffic, recently the cellular operators are deploying more and more base stations (BSs) and their daily energy cost amounts to a large portion of the operational expenditure (OPEX). For example, China Mobile owns around 920 thousand BSs by 2011 and the total energy cost per year is almost 3 billion US dollars, given that the annual cost for each BS is about 3 thousand US dollars \cite{Hasan}. Therefore, many cellular operators want to reduce their energy costs by employing new cost-saving solutions \cite{Hasan,WuJinsong,HanAnsari2014,FRYu2012}, which in general manage either the energy supply or the communication demand of cellular networks.

On the supply side, one commonly adopted solution is to use energy harvesting devices (e.g., solar panels and wind turbines) at cellular BSs, which can harvest the cheap and clean renewable energy to reduce or even substitute the energy purchased from the grid \cite{HanAnsari2014}. However, since the renewable energy is often randomly distributed in both time and space, different BSs are hard to solely use their individually harvested energy to power their operations. As a result, the power grid is still needed to provide reliable energy to BSs. Besides serving as a reliable energy supply, power grid also provides new opportunities for the BSs' cost-saving with its ongoing paradigm shift from traditional grid to smart grid. Unlike traditional grid, which uses one-way energy flow to deliver power from central generators to electricity users, smart grid deploys smart meters at end users to enable both two-way information and energy flows between the grid and end users \cite{XueSmartGird,SaadSPM}. The two-way energy flow in smart grid motivates a new idea of {\it energy cooperation} in cellular networks, as we will elaborate later in this article, which allows the BSs to trade and share their unevenly harvested renewable energy through the smart grid to support the non-uniform wireless traffic in a cost-effective way.

\begin{figure}
\centering
 \epsfxsize=1\linewidth
    \includegraphics[width=12cm]{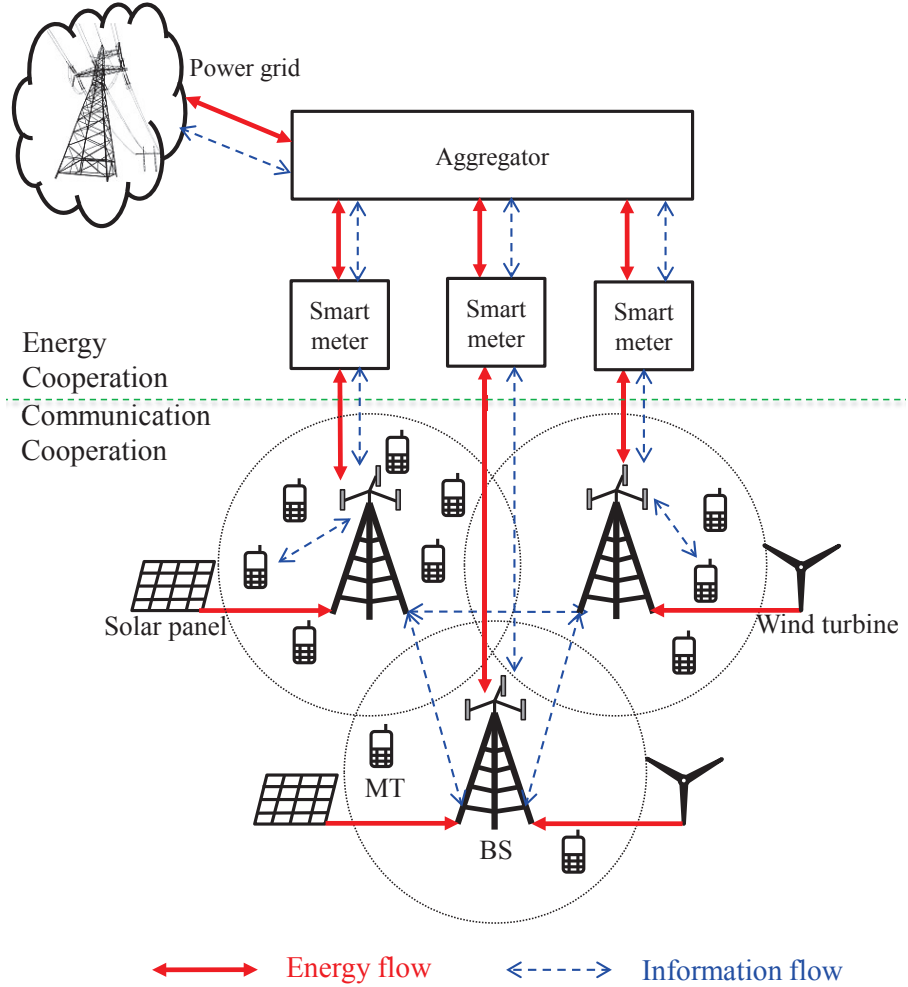}
\caption{A general model of cellular networks with energy and communication cooperation among BSs.
} \label{fig:2}
\end{figure}

On the demand side, various techniques have been proposed in cellular networks across different layers of communication protocols for reducing the energy consumption \cite{Hasan}. Among them, communication cooperation (e.g., traffic loading \cite{Niu:CellZooming}, spectrum sharing \cite{Goldsmith:Spectrum} and coordinated multi-point (CoMP) \cite{Gesbert2010}) is particularly appealing, which allows the BSs to share the wireless resources and shift the traffic loads with each other for energy-saving. However, the introduction of renewable energy at BSs imposes new challenges on the existing communication cooperation design: the conventional energy-saving design may not be cost-effective any longer. This is due to the fact that the renewable energy (though unreliable in supplying) is in general much cheaper than the energy purchased from the grid and therefore BSs should maximally use it to save cost, whereas under the energy-saving design the harvested renewable energy at BSs may not be efficiently utilized when serving the time- and space- varying wireless traffic. To overcome this problem, it is desirable to design new {\it cost-aware communication cooperation} approaches, by taking into account the cost differences between renewable and conventional energy.

In this article, we first overview the recent advances in energy cooperation and cost-aware communication cooperation. Fig. \ref{fig:2} illustrates the general energy and communication cooperation concept for cellular networks at both the energy supply layer and the communication demand layer, respectively. Then we propose a new {\it joint energy and communication cooperation} to exploit both benefits.  Specifically, we present the following three approaches.
\begin{itemize}
  \item {\bf Approach I: energy cooperation on the supply side}. Cellular systems or BSs use the two-way energy flow in smart grid to trade or share renewable energy, by taking the energy demands for communications as given.
\item  {\bf Approach II: communication cooperation on the demand side}. Cellular systems or BSs  perform cost-aware communication cooperation to share wireless resources and reshape wireless load over space and time, by taking the energy supply (renewable and/or conventional) as given.
  \item {\bf Approach III: joint energy and communication cooperation on both sides}. Cellular systems or BSs  jointly cooperate on both the supply and demand sides to maximally reduce their total energy cost.
\end{itemize}

In the rest of this article, we first introduce the energy supply and demand models of cellular systems. Then we present latest energy, communication, and joint cooperation approaches, respectively. Finally, we point out several future research directions and conclude this article.

\section{Energy Supply and Demand of Cellular Systems}

In this section, we introduce the energy supply and demand models for cellular systems. The models will be used to motivate various cooperation schemes introduced later. {\color{black}{For notational convenience, in this article we  focus on one particular time slot and thus skip the time index of variables (such as the energy demand and supply at BSs as well as the energy prices), as will be shown later. Note that these variables can vary over time in practice. In addition, we normalize the length of time slot into unity, and thus use the terms energy and power interchangeably throughout this article.}}

\begin{figure}
\centering
 \epsfxsize=1\linewidth
    \includegraphics[width=10cm]{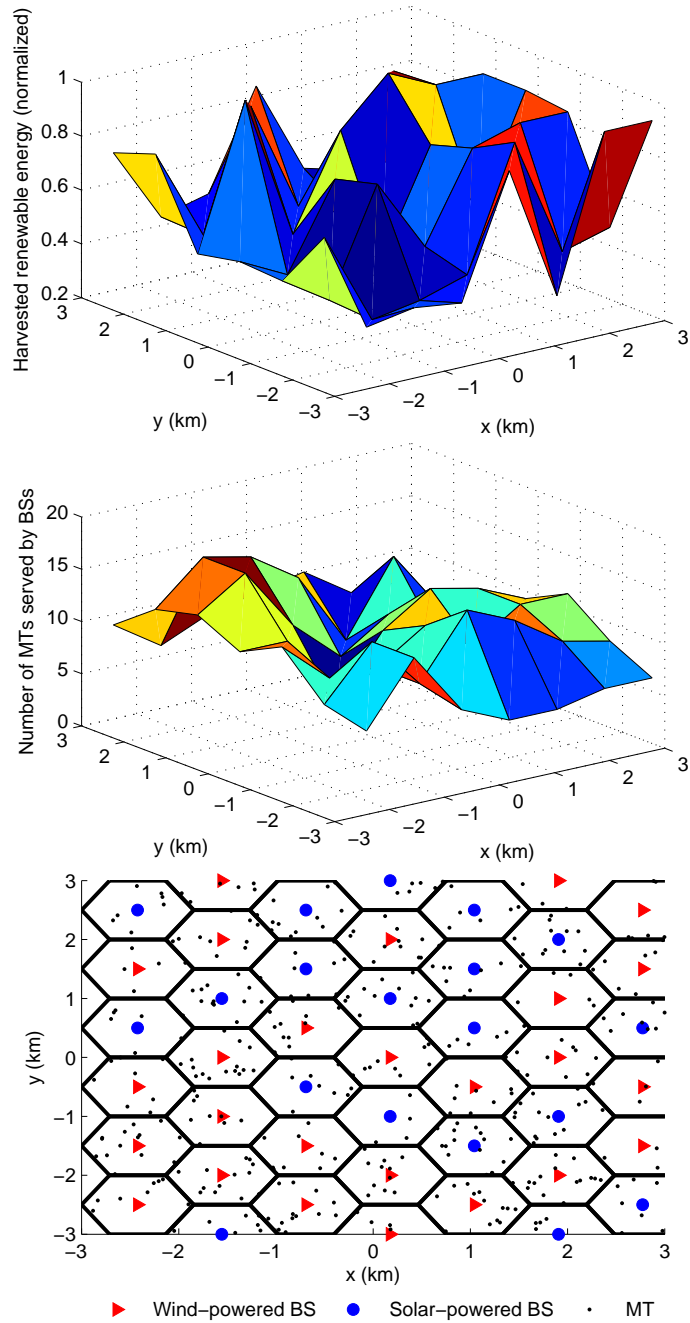}
\caption{An example of a cellular system with BSs having spatially distributed traffic load and harvested energy at a given time instance. It is assumed that the (normalized) energy harvesting capacity of all solar-powered BSs is 1, and that of all wind-powered BSs is 0.5.} \label{fig:1:example}
\end{figure}

We consider a single cellular system with $N > 1$ BSs, in which each BS $i$ is deployed with a renewable energy harvesting device that has a harvesting rate $E_i \ge 0, i=1,\ldots,N$. The value of $E_i$ at a given time instance depends on the type of the renewable energy source (e.g., solar or wind), the harvesting capacity of the device (e.g., size of the solar panel), and the weather condition at that location. As shown in the upper sub-figure of Fig. \ref{fig:1:example}, the $E_i$'s are generally different among BSs at different locations.

On the demand side, the power consumption of each cellular BS $i$, denoted by $Q_i \ge 0$, is composed of two main parts: the dynamic power consumption related to the transmission and reception of wireless signals for serving the mobile terminals (MTs), and the constant power consumption (e.g., at the circuits and air conditioners) for maintaining other operations. In reality, the value of $Q_i$ varies according to the traffic load over the service coverage area of BS $i$. Due to MTs' mobility across cells and their time-varying service requests, the traffic loads (and thus the $Q_i$'s) are different among BSs and change over time, as shown in the middle sub-figure of Fig. \ref{fig:1:example}.

By combining the supply and demand sides, we denote the {\it net load} at BS $i$ as $\delta_i = Q_i - E_i$, where $\delta_i > 0$ shows the deficit status of renewable energy and $\delta_i < 0$ indicates the energy surplus status. Since $Q_i$'s and $E_i$'s are usually independent (see Fig. \ref{fig:1:example}), it is likely that some BSs are short of renewable energy to match demand (i.e., $\delta_i > 0$), while the other BSs are adequate in renewable energy (i.e., $\delta_i < 0$). Such a geographical diversity in net load requires some BSs to purchase energy from the grid (e.g., a $\delta_i$ amount of energy purchase for BS $i$ with $\delta_i>0$) but the other BSs to waste the extra renewable energy (i.e., a $|\delta_j|$ amount of energy waste for BS $j$ with $\delta_j<0$).{\footnote{\color{black}One possible solution to reduce such renewable energy waste is to equip storage devices at BSs to store extra energy for future use. However, currently the storage devices are expensive and capacity-limited, and commercial BSs often do not equip such devices for dynamic energy management.}} Overall, the total purchased energy amount from the grid by all the $N$ BSs is the total renewable energy deficit, denoted by $\Delta_+\triangleq \sum_{i=1}^{N}  [\delta_i]^+ \ge 0$ with $[x]^+ \triangleq \max(x,0)$, while the total wasted renewable energy amount by them is the total renewable energy surplus, given by $\Delta_-\triangleq - \sum_{j=1}^{N}  [\delta_j]^- \ge 0$ with $[x]^- \triangleq \min(x,0)$. By denoting the price for BSs to purchase energy from the grid as $\pi > 0$,  then the total energy cost of the cellular system is
\begin{align}\label{eqn:1}
C_1 =  \pi \Delta_+,
\end{align}
which is independent of $\Delta_-$. This fact motivates us to use the wasted renewable energy surplus ($\Delta_-$) to compensate the deficit ($\Delta_+$) for cost-saving. To this end, we will implement the energy and communication cooperation on the supply and demand sides, respectively, to reschedule and balance $E_i$'s and $Q_i$'s.

\section{Energy Cooperation}\label{sec:energy_coop}

Energy cooperation is a cost-saving approach on the supply side, in which the cellular BSs are allowed to employ two-way energy trading or sharing for better utilizing their otherwise wasted renewable energy surplus ($\Delta_-$). {\color{black}Although the idea of energy cooperation has been mentioned in smart grid for microgrids' energy trading \cite{SaadSPM}, it is new to cellular networks. Particularly, since it is too complex for the grid to directly control a large number of BSs, the energy trading and sharing in cellular networks should be enabled by using aggregators \cite{Gkatzikis2013} (see the upper energy cooperation layer in Fig. \ref{fig:2}).} With aggregators, we can cluster BSs into a finite number of groups and an aggregator can serve as an intermediary party to control each group of BSs for the grid, thus helping realize the two-way energy flow between the grid and BS groups. The implementation of energy cooperation is not difficult in smart grid: it only requires the two-way energy flow and aggregators, and does not change the existing infrastructure of cellular networks.


\subsection{Aggregator-Assisted Energy Trading}\label{sec:BS2Grid}

Aggregator-assisted energy trading is an energy cooperation scheme in which the aggregator performs two-way energy trading with the BSs by deciding buying and selling prices. In this scheme, the BSs adequate in renewable energy can sell their extra energy to the aggregator, from which the selling revenue can be gained to compensate the total energy cost; at the same time, the other BSs short of renewable energy can obtain such cheap energy from the aggregator at a lower price than the regular price $\pi$ to purchase from the grid directly. As the coordinator in this trading market, the aggregator can also obtain some revenue by properly deciding the energy selling and buying prices. {\color{black}Here, the energy selling and buying at each BS is managed with the help of  the smart meter in real time, which can decide the sold/purchased energy amount at any time slot {\it a priori} based on the energy harvesting rates, the power demand and the energy prices. It does not strictly require BSs to deploy energy storage devices.}

Let $\pi_{\rm buy} > 0$ and $\pi_{\rm sell}> 0$ denote the unit prices for each BS to buy and sell energy from and to the aggregator, respectively.{\footnote{\color{black}The energy prices may vary over time based on the time-varying relationship between aggregate energy demand and supply at the aggregator. Depending on information and energy exchange frequencies, the aggregator can decide such prices either day-ahead or in real time.}} Here, $\pi_{\rm sell}<\pi_{\rm buy}$ holds to avoid the trivial case where a BS can benefit by reselling its bought energy from the aggregator; and $\pi_{\rm buy}<\pi$ is also true, since otherwise all BSs short of energy will buy cheaper energy from the grid directly. With the two-way energy trading, the BSs adequate in renewable energy will sell their total $\Delta_-$ amount of energy surplus to the aggregator at the price $\pi_{\rm sell}$, and accordingly an energy quota is set by the aggregator as $\Delta_-$. The BSs short of renewable energy will first purchase a $\min(\Delta_+, \Delta_-)$ amount of cheap energy from aggregator at the price $\pi_{\rm buy}$ (with the quota limitation of $\Delta_-$) to maximally use this resource, and (if not enough) will buy a $\Delta_+ - \min(\Delta_+, \Delta_-)$ amount from the grid at the price $\pi$. Depending on the relationship between $\Delta_+$ and $\Delta_-$, the total cost of all the $N$ BSs is
\begin{eqnarray} \label{eqn:2}
C_2=\left\{\begin{array}{ll} \pi_{\rm buy}\Delta_+ - \pi_{\rm sell}\Delta_-, & {\rm if}~\Delta_+ \le \Delta_- \\ \pi_{\rm buy}\Delta_-  + \pi \left(\Delta_+ - \Delta_-\right) - \pi_{\rm sell}\Delta_-, &{\rm if}~ \Delta_+ > \Delta_-. \end{array} \right.
\end{eqnarray}
Note that $C_2$ can be even negative, which is the case when $\Delta_-$ is sufficiently larger than $\Delta_+$ such that $\pi_{\rm buy}\Delta_+ < \pi_{\rm sell}\Delta_-$. By comparing  (\ref{eqn:1}) and (\ref{eqn:2}), it follows that $C_{2} \le C_1$.

\subsection{Aggregator-Assisted Energy Sharing}\label{sec:BS2BS}

Aggregator-assisted energy sharing is another energy cooperation scheme that allows BSs in a BS group to mutually negotiate and share renewable energy {\color{black}by simultaneously injecting and drawing energy to and from the aggregator}, respectively. By matching the local renewable energy deficit (positive $\delta_i$'s) and surplus (negative $\delta_i$'s) between any two BSs, this scheme helps the group of BSs reduce their aggregate renewable energy deficit. The practical implementation of the energy sharing requires this group of BSs to sign a {\it contract} with the aggregator at a contract fee that motivates the aggregator to support the energy sharing. Note that the contract still requires BSs to commit to not interfere the common operation of the aggregator, by equating their total injected energy (into the aggregator) to their total drawn energy (from the aggregator) at any given time instance.{\footnote{The aggregator can alternatively provide a long-term contract to the BSs, such that the BSs only need to ensure the energy sharing balance over a longer time period (e.g., one day or even a month), thus offering more flexibility for the BSs' energy sharing. In this case, however, a higher contract fee may be required by the aggregator due to the short-term interference to the aggregator.}} Compared to aggregator-assisted energy trading scheme, the aggregator does not need to be actively involved in this energy sharing scheme and endures limited coordination complexity.

Specifically, suppose that BS $i$ wants to transfer an $e_{ij} \ge 0$ amount of energy to BS $j$, $i\neq j$. This is accomplished at an appointed time by BS $i$ injecting an $e_{ij}$ amount of energy into the aggregator, and at the same time BS $j$ drawing the same $e_{ij}$ amount from the aggregator.{\footnote{\color{black}The energy transfer between two BSs through the aggregator may lead to a certain amount of energy loss \cite{XuGuoZhangGC2013}. Since the loss is very small (e.g., less than 5\% of the transferred energy), its impact is neglected here though our scheme also applies to that case.}} Thanks to the mutual sharing of $e_{ij}$'s among the $N$ BSs, the total energy deficit $\Delta_+$ and surplus $\Delta_-$ can be effectively matched. When $\Delta_+ \le \Delta_-$, the $N$ BSs can maintain their operation without purchasing any energy from the grid; otherwise, a total $\Delta_+-\Delta_-$ amount of energy should be purchased from the grid at the price $\pi$. By denoting the contract fee to the aggregator as $\bar C$, the total cost of all the $N$ BSs is given by
\begin{eqnarray} \label{eqn:3}
C_3=\left\{\begin{array}{ll}
\bar C, &{\rm if}~ \Delta_+ \le \Delta_- \\ \pi \left(\Delta_+-\Delta_-\right) + \bar C, & {\rm if}~\Delta_+ > \Delta_-. \end{array} \right.
\end{eqnarray}
By comparing (\ref{eqn:3}) and (\ref{eqn:1}), it generally follows that $C_3 \le C_1$, i.e., the total energy cost is reduced, as long as $\bar C$ is sufficiently small.

The aggregator-assisted energy trading and sharing schemes may have different cost-saving performances depending on the energy buying and selling prices in the former scheme, and the contract fee in the latter one, both of which incentivize the aggregator to help.


\section{Communication Cooperation}

Communication cooperation refers to a cost-saving approach on the demand side that exploits the broadcast nature of wireless channels and uses wireless resource sharing to reshape BSs' wireless load and energy consumption. {\color{black}Different from conventional communication cooperation (e.g., \cite{Niu:CellZooming,Goldsmith:Spectrum,Gesbert2010}) aiming to maximize data throughput or minimize energy consumption, the communication cooperation of our interest here seeks to minimize the total energy cost by optimally utilizing both the cheap renewable energy and reliable on-grid energy. In the so-called cost-aware communication cooperation, the rescheduling of BSs' traffic load and energy consumption should follow their given renewable energy supply, such that the renewable energy can be maximally used to support the quality of service (QoS) requirements of the MTs, and the on-grid energy purchase is thus minimized.} To implement this approach, BSs should share with each other the communication information (e.g., channel state information and QoS requirements of MTs) and the energy information (e.g., the energy harvesting rates) through the backhaul links connecting them, as shown in Fig. \ref{fig:2}. This may require the cellular operator to install new infrastructures (e.g., high-capacity and low-latency backhaul links) and/or coordinate and standardize the communication protocols. It may incur more implementation complexity than the energy cooperation in Section \ref{sec:energy_coop}.

\begin{figure}
\centering
 \epsfxsize=1\linewidth
    \includegraphics[width=14cm]{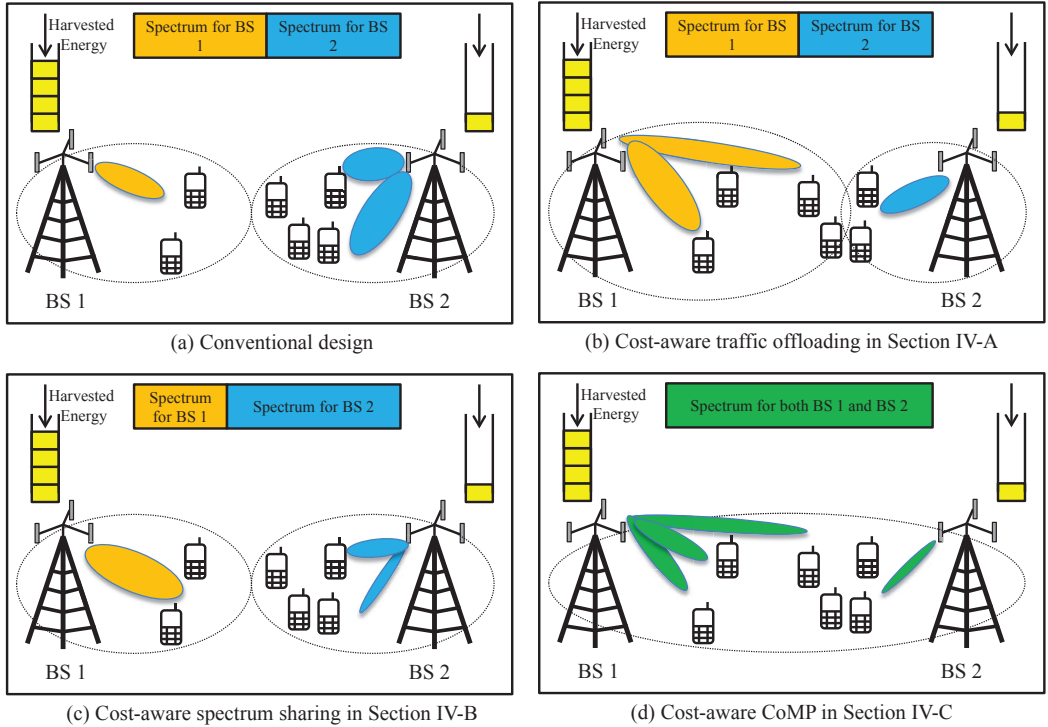}
\caption{An example of different communication cooperation designs in a simple cellular network with two BSs.} \label{fig:3}
\end{figure}

In this section, we discuss three different cost-aware communication cooperation schemes, namely, traffic offloading \cite{Niu:CellZooming}, spectrum sharing \cite{Goldsmith:Spectrum}, and coordinated multi-point (CoMP) \cite{Gesbert2010}, which are implemented in different time scales. For the purpose of illustration, we consider a simple cellular system setup with two BSs as shown in Fig. \ref{fig:3}(a), in which BS 1 has sufficient harvested renewable energy and light traffic load (serving 2 MTs), thus having the net load $\delta_1 < 0$; while BS 2 has insufficient renewable energy and heavy traffic load (serving 4 MTs), leading to the net load $\delta_2 > 0$. {\color{black}The corresponding spectrum and power allocation and user-BS association for the two BSs are shown in Fig. \ref{fig:3}(b-d) for different cost-aware communication cooperation schemes.}

\subsection{Cost-Aware Traffic Offloading}

Traffic offloading is traditionally designed to shift the traffic load (or served MTs) of heavily loaded BSs to the lightly loaded ones, for the purpose of avoiding the traffic congestion and improving the QoS of the MTs. Differently, the cost-aware traffic offloading here focuses on the new issue of energy cost reduction, i.e., BSs short of renewable energy can offload their MTs to neighboring BSs with abundant renewable energy (even if they have more or similar traffic loads), thus reducing the whole energy drawn from the gird to save cost. As shown in the example of Fig. \ref{fig:3}(b), it is cost-effective for BS 2 to offload 2 MTs (at its cell edge)  to BS 1, such that the renewable energy at BS 1 is better utilized. Traffic offloading is often employed in a time scale of several seconds.

\subsection{Cost-Aware Spectrum Sharing}\label{sec:spectrum_sharing}

Besides energy, spectrum is another scare resource in cellular networks and spectrum sharing has been considered as a solution to improve the spectrum utilization efficiency \cite{Goldsmith:Spectrum}. Different from the conventional spectrum sharing, the cost-aware spectrum sharing is based on the fact that the energy and spectrum resources can partially substitute each other to support the wireless transmission, and sharing spectrum to a BS short of energy can better save the energy cost of that BS.{\footnote{Considering a point-to-point additive white Gaussian noise (AWGN) channel, the relationship between the transmit power $P \ge 0$ and the bandwidth $W\ge 0$ is given by $P = N_0W\left(2^{r/W}-1\right)$, where $r$ and $N_0$ denote the transmission rate and the noise power spectral density at the receiver, respectively. In this case, as the system bandwidth $W$ increases, the transmit power $P$ will decrease, and vice versa.} As shown in the example of Fig. \ref{fig:3}(c), BS 1 shares part of its available spectrum to BS 2. Under the same QoS requirements of MTs, BS 2 can decrease its transmission power purchased from the grid, while BS 1 uses more renewable energy for its transmission. Hence, the total cost is reduced. The implementation of spectrum sharing requires the BSs (and perhaps even the MTs) to have the capability of aggregating different frequency bands in transmission and reception, e.g., with the advanced carrier aggregation technique. Spectrum sharing can be realized in a time scale of minutes.

\subsection{Cost-Aware Coordinated Multi-Point (CoMP)}\label{sec:comp}

Traditionally, CoMP is considered as a technique to improve the spectral efficiency in cellular networks, by which BSs can implement coordinated baseband signal processing to cooperatively serve multiple MTs over the same time-frequency resources, transforming the harmful inter-cell interference (ICI) into useful information signals \cite{Gesbert2010}. Differently, the cost-aware CoMP is motivated by the following observation: since different BSs can cooperatively send information signals to the MTs (in the downlink), their transmission power can be compensated by each other for satisfying the QoS requirements at MTs. Therefore, by adaptively adjusting the BSs' transmit signals, the cost-aware CoMP helps match the BSs' transmission power with their harvested renewable energy, thus minimizing the total energy drawn from the grid to save cost. For example, in Fig. \ref{fig:3}(d), BS 1 adequate in renewable energy should use a high transmission power for providing strong wireless signals to the MTs, while BS 2 short of renewable energy should transmit at a low power level in their CoMP transmission.

CoMP should be performed at a symbol or frame level in the time scale of microseconds/milliseconds, which is more complex than the aforementioned traffic offloading and spectrum sharing schemes, but can achieve higher cost-saving (as will be shown in the case study later). In practical cost-aware cellular networks, the three communication cooperation schemes can be employed depending on the trade-off between cost-saving and implementation complexity.


\section{Joint Energy and Communication Cooperation}

Joint energy and communication cooperation can maximally save cost by applying both the energy cooperation on the supply side and communication cooperation on the demand side. {\color{black}To realize the joint operation, the BSs should share the energy information by using the two-way information flow supported by the smart gird (through the smart meters), and also exchange the communication information through their backhaul connections (see Fig. \ref{fig:2}). Here, the exact required information sharing among BSs depends on the specific energy and communication cooperation schemes employed.

The joint energy and communication cooperation is more complex than energy or communication cooperation only, due to the implementation complexity for solving the cost minimization problem by optimizing both the supply (e.g., energy trading/sharing among BSs) and demand (e.g., spectrum and power allocations at BSs) sides, as well as the signaling overhead for sharing both the energy and communication information among BSs. The complexity increases significantly as the network size or the number of BSs becomes large. One potential solution to resolve this problem is to dynamically group the huge number of BSs into different BS clusters, where BSs within each cluster can implement the joint cooperation in a centralized manner, and different clusters can perform limited coordination in a decentralized way. In this section, we focus our study on the joint cooperation among a limited number of BSs in one single cluster.}

As there are two energy cooperation schemes (aggregator-assisted energy trading and energy sharing) in Section III and three communication cooperation schemes (traffic offloading, spectrum sharing, and CoMP) in Section IV, there are totally six combinations of joint cooperation designs. In this section, we focus on three specific schemes: a joint energy and spectrum sharing design and two joint energy cooperation and CoMP designs. The ideas can be similarly extended to the other three combinations.

\subsection{Joint Energy and Spectrum Sharing}

The joint energy and spectrum sharing \cite{GuoTCOM} is a scheme that allows neighboring BSs to share energy and spectrum with each other through the  aggregator-assisted energy sharing in Section \ref{sec:BS2BS} and the spectrum sharing in Section \ref{sec:spectrum_sharing}, respectively. {\color{black}In this scheme, the BSs share their energy harvesting rates, energy prices, available bandwidth, and channel state information (e.g., channel gains), as well as the QoS requirements of MTs among each other.} Accordingly, the BSs exchange energy and spectrum to take advantage of resource complementarity.

Building upon the spectrum sharing in Fig. \ref{fig:3}(c), \cite{GuoTCOM} considered joint energy and spectrum sharing between two BSs to minimize their total energy cost, while ensuring the QoS requirements for all the MTs. It is shown that at the optimality, it is possible that one BS adequate in both energy and spectrum shares these two resources to the other (in unidirectional cooperation), or one BS exchanges its energy for spectrum with the other (in bidirectional cooperation).

\subsection{Joint Energy Cooperation and CoMP}

In this scheme, different BSs implement CoMP-based transmission/reception in Section \ref{sec:comp} to serve one or more MTs over the same time frequency resources, and at the same time perform aggregator-assisted energy trading in Section \ref{sec:BS2Grid} or aggregator-assisted energy sharing in Section \ref{sec:BS2BS}. {\color{black}To implement so, the BSs need to share their energy harvesting rates, energy prices, and instantaneous channel state information (both channel gains and phases), as well as the QoS requirements of MTs among each other.} Based on different types of energy cooperation schemes employed, \cite{XuZhangGC2014} and \cite{XuGuoZhangGC2013} studied two different joint energy cooperation and CoMP designs for the downlink transmission.

First, when aggregator-assisted energy trading in Section \ref{sec:BS2Grid} is implemented among BSs, \cite{XuZhangGC2014} proposed to jointly optimize the BSs' cooperative transmit beamforming in CoMP based communication and their two-way energy trading with the aggregator, so as to minimize their total energy cost. It is shown that by exploiting the non-uniform harvested renewable energy $E_i$'s over different BSs and the difference between the energy buying and selling prices $\pi_{\rm buy}$ and $\pi_{\rm sell}$, the joint energy trading and CoMP optimization achieves a significant cost reduction, as compared to the design that separately optimizes the CoMP based communication and the energy trading.


Next, when aggregator-assisted energy sharing in Section \ref{sec:BS2BS} is adopted, \cite{XuGuoZhangGC2013} considered to use it to enable a new purely renewable-powered cellular system, in which the BSs do not purchase any energy from the grid to minimize the cost, but use the harvested renewable energy together with the energy sharing to maintain their operations. By taking into account the possible energy loss during the energy sharing, \cite{XuGuoZhangGC2013} maximizes the weighted sum-rate for all served MTs in one particular CoMP cluster by jointly optimizing the cooperative BSs' zero-forcing beamforming design and their shared energy amounts among each other.

\subsection{A Case Study}
\begin{figure}
\centering
 \epsfxsize=1\linewidth
    \includegraphics[width=12cm]{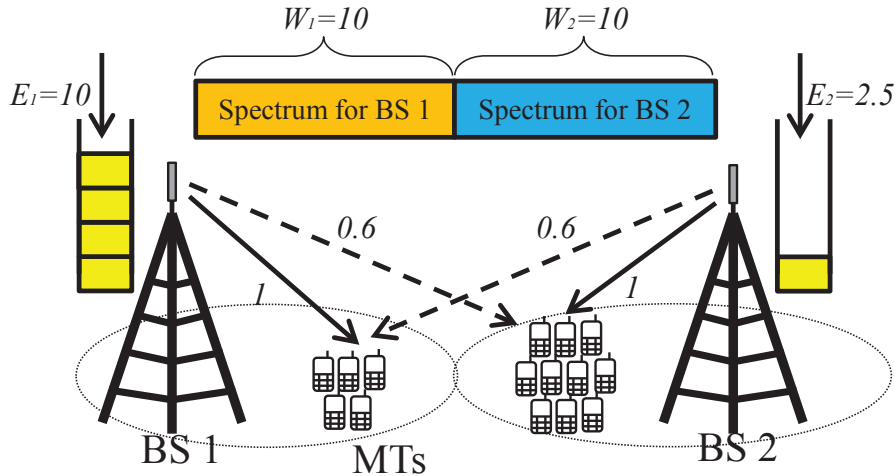}
\caption{The case study model for comparing energy cooperation, communication cooperation, and joint cooperation.} \label{fig:1}
\end{figure}
Now we present a case study to compare the energy cooperation in Section III, the cost-aware spectrum sharing and CoMP in Section IV, and the three joint energy and communication cooperation schemes proposed in this section. {\color{black}Also, we consider the conventional design without energy or communication cooperation as the performance benchmark, where each BS first individually minimizes its energy consumption on the demand side while ensuring the QoS requirements at MTs, and then (if the energy demand exceeds the renewable energy supply) purchases the additional energy from the grid. For the purpose of illustration, as shown in Fig. 4, we consider the downlink of a cellular system with two single-antenna BSs (i.e., BS 1 and BS 2) each applying orthogonal frequency-division multiple access (OFDMA) to serve $K_1 = 5$ and $K_2 = 15$ single-antenna MTs (denoted by the MT sets $\mathcal K_1$ and $\mathcal K_2$), respectively. Each BS uses an orthogonal frequency band with the same bandwidth  ($W_1 = W_2 = 10$). For simplicity, we randomly generate the channels based on the independent and identically distributed (i.i.d.) Rayleigh fading with the average channel powers from each BS to its own associated MTs (i.e., from BS 1 to any MT in $\mathcal K_1$ and from BS 2 to any MT in $\mathcal K_2$) being 1, and that from each BS to the other BS's associated MTs (i.e., from BS 1 to any MT in $\mathcal K_2$ and from BS 2 to any MT in $\mathcal K_1$) being 0.6. We set the noise power spectral density at each MT to be 1, and the QoS requirement of each MT to be a minimum data rate $1$. On the demand side, we set the power consumptions $Q_1$ and $Q_2$ at the two BSs as their transmission power only; on the supply side, we set their harvested renewable energy as $E_1= 10$ and $E_2= 2.5$, respectively, and their energy buying price from the grid as $\pi =1$. Additionally, for the aggregator-assisted energy trading, the BSs' energy buying and selling prices from and to the aggregator are $\pi_{\rm buy}  = 0.5$ and $\pi_{\rm sell} = 0.4$, respectively; and for the aggregator-assisted energy sharing, the contract fee paid to the aggregator is $\bar C = 0.1$. Furthermore, in each scheme, the BSs employ the equal bandwidth allocation among MTs and there is only one MT served in each sub-band. Note that all units are normalized for simplicity here.}

Based on the above setting, we summarize the results in Table \ref{table2}, from which we have the following observations.

\definecolor{Mahogany}{RGB}{0, 100, 0}

\begin{table}[!t]\scriptsize
\caption{Energy Cost Performance Comparison}
\label{table2} \centering
\begin{tabular}{|p{1.75in}|p{0.8in}|p{0.8in}|p{0.7in}|p{0.7in}|p{0.7in}|}
\hline
&{BS 1's renewable energy supply}&{BS 2 's renewable energy supply}&{BS 1's energy consumption}&{BS 2's energy consumption}&{Total energy cost}\\
\hline
{Conventional design without energy or communication cooperation }& $10$&$2.5$&{$4.14$}&	{$18.28$}&	{${\color{black}\bf 15.78}$}\\
\hline
{Approach I: energy cooperation via aggregator-assisted energy trading}&${\color{red}4.14}$&${\color{red}8.36}$&{$4.14$}&{$18.28$}&{${\color{red}\bf 10.51}$}\\
\hline
{Approach  I: energy cooperation via aggregator-assisted energy sharing}&${\color{red}4.14}$&${\color{red}8.36}$&{ $4.14$}&	{$18.28$}&{${\color{red}\bf 10.03}$}\\
\hline
{Approach  II: communication cooperation via spectrum sharing}&$10	$&$ 2.5$&{ ${\color{blue}10.00}$}	&{	${\color{blue} 14.04}$	}&{ ${\color{blue}\bf  11.54}$}\\
\hline
{Approach  II: communication cooperation via CoMP}&$10$&$2.5$&{ ${\color{blue}10.00}$}&	${\color{blue}3.75}$	&{ ${\color{blue}\bf  1.25}$}\\
\hline
{Approach  III: joint energy and spectrum sharing}&${\color{red}5.00}$&${\color{red}7.50}$& { ${\color{blue}5.00}$	}&{${\color{blue}15.00}$}& {${\bf \color{Mahogany}7.60}$}\\
\hline
{Approach  III: joint aggregator-assisted energy trading and CoMP}&${\color{red}6.87}$&${\color{red}5.62}$& { ${\color{blue}6.87}$	}&{${\color{blue}5.77}$}& {${\bf \color{Mahogany}0.46}$}\\
\hline
{Approach III: joint aggregator-assisted energy sharing and CoMP}&${\color{red} 5.47}$&${\color{red}7.03
}$& { ${\color{blue} 5.47}$	}&{${\color{blue}7.03
}$}& {${\bf \color{Mahogany}0.10}$}\\
\hline\end{tabular}
\end{table}

\begin{itemize}
  \item For the conventional design, the two BSs' energy demands for communications are computed to be $Q_1 =  4.14$ and $Q_2 = 18.28$. Their total energy cost is $15.78$.
  \item For both energy cooperation approaches, it is observed that the renewable energy supplies at BS 1 and BS 2 are rescheduled to be $4.14$ and $8.36$, respectively, by BS 1 trading or sharing its excessive renewable energy to BS 2 through the aggregator. Since the new renewable energy supplies better match with the given energy consumptions at the two BSs, their total energy cost reduces to $10.51$ and $10.03$, respectively, where the different cost reductions are related to the different service fees charged by the aggregator.


  \item Regarding communication cooperation, it is observed that the energy consumptions at the two BSs are respectively changed to $Q_1 = 10.00$ and $Q_2 = 14.04$ for the spectrum sharing scheme, and $Q_1 = 10.00$ and $Q_2 =  3.75$ for the CoMP scheme. Compared to the conventional design, communication cooperation increases the transmission power at BS 1 to partially substitute that at BS 2 (together with certain wireless resource sharing) while satisfying the MTs' QoS requirements. The transmission power adaptation matches and better uses the given cheap renewable energy supplies at the two BSs. Consequently, the resulting total energy costs reduce to $11.54$ and $1.25$, respectively.




  \item For joint energy and communication cooperation, it is observed that by exploiting both supply- and demand-side management, each joint scheme outperforms the corresponding schemes with only energy or communication cooperation. For instance, the total energy cost of joint energy and spectrum sharing ($7.6$) is less than that of aggregator-assisted energy sharing only ($10.03$) and spectrum sharing only ($11.54$). Furthermore, it is observed that the two joint energy cooperation and CoMP designs achieve the lowest total energy cost ($0.46$ and $0.10$, respectively) and outperform all the other schemes. Therefore, the joint energy cooperation and CoMP design is promising for maximum cost-saving.

\end{itemize}


\section{Extensions and Future Directions}

Despite the aforementioned studies on energy and communication cooperation, there remain a lot of interesting topics unaddressed. We list several of them as follows for future study.


{\color{black}Practically, energy harvesting rates in general change slowly as compared to wireless channel and traffic load variations, and as a consequence, the time scale of implementing energy cooperation is normally longer than that of communication cooperation. However, joint energy and communication cooperation in this article requires the energy cooperation to be realized in the same time scale as the communication cooperation, thus needing more frequent decision making at BSs and higher implementation complexity at smart meters. To overcome this issue, it is promising to consider the multi-time-scale implementation of joint energy and communication cooperation, e.g., by employing a two-layer decision making with energy cooperation in longer time scale and communications cooperation in shorter time scale, so as to balance the tradeoff between the cost-saving performance and the implementation complexity.}

So far, we have focused on a single cellular system or multiple systems belonging to the same entity, aiming to minimize the total energy cost. In practice, however, multiple self-interested systems (owned by different operators) can coexist or co-locate and it is interesting to study their energy and/or communication cooperation. Unlike energy trading in Section III, on the energy supply side, more than one aggregators may be needed to facilitate trading across different BS groups. As for the mutual energy sharing scheme, one selfish system may want to sell (buy) renewable energy to (from) the other system at a high (low) price. On the communication demand side, it is happening that some systems (e.g., Verizon and T-Mobile) are sharing spectrum in long-term. Yet how to enable communication cooperation in short-term (as in Section IV) requires inter-system communication compatibility and more coordination. Moreover, to establish joint energy and communication cooperation, cellular systems may seek for the advantage of resource complementarity. For example, in a preliminary study \cite{GuoTCOM}, it is shown that one system adequate in spectrum is willing to cooperate with another adequate in energy, since both systems can efficiently reduce their individual costs by exchanging spectrum and energy with each other. Overall, cooperation mechanism design is required to motivate or strengthen inter-system joint cooperation to a win-win situation for all systems involved.

Besides cellular networks, it is also appealing for heterogeneous communication networks (e.g., WiFi and small cells) to cooperate and reduce overall energy cost. Offloading a mobile user from a macrocell to a small cell saves energy, and better utilizes the wireline backhaul resource to expand the limited wireless spectrum. Yet these networks are different in service coverage, operated spectrum, and even energy harvesting availability (difficult indoor), and their joint energy and communication cooperation becomes more complicated than our design in Section V. For example, scalability could be a problem and one possible solution is to decompose the whole heterogeneous network into a number of micro-networks as in \cite{XueSmartGird} with cooperation in each.

Up to now, we have focused on the case without the use of energy storage at BSs due to the cost consideration. With the advancement of battery technologies, we envision that energy storage may be employed in the future BSs and it is promising to study the energy and communication cooperation jointly with the storage management. In principle, the storage devices handle the renewable energy fluctuations at BSs to match the energy demand variations over time, while the energy and communication cooperation approaches do that over space. Therefore, the two approaches can be good complementarities. {\color{black}Nevertheless, such joint time and space domain optimization problems are very challenging to solve, since any present decisions made by BSs would affect their storage status and traffic loads served in the future.} As an initial study, \cite{Chia2013,TJLim2014} have considered the joint energy cooperation and storage management problem for minimizing the total energy cost in a simplified cellular system with given energy demand at BSs.


%
%
%


\section{Conclusion}

This article provides an overview on novel energy and communication cooperation approaches for energy cost-saving in cellular networks powered by renewable energy sources and smart grid. These approaches use both the two-way energy flow in smart grid and the communication cooperation in cellular networks to reshape the non-uniform energy supplies and energy demands over the cellular networks for cost-saving. It is our hope that these new approaches can bring new insights on the energy demand management in smart grid by considering the unique properties of the cellular networks' communication demand, and also on the wireless resource allocation in cellular networks by taking into account the new characteristics of the emerging renewable and smart grid energy supply.





\begin{thebibliography}{1}
\bibliographystyle{IEEEbib}
\bibitem{Hasan}
Z. Hasan, H. Boostanimehr, and V. Bhargava, ``Green cellular networks: a survey, some research issues and challenges,'' {\it IEEE Commun.
Surveys \& Tutorials}, vol. 13, no. 4, pp. 524-540, 2011.

\bibitem{WuJinsong}
J. Wu, S. Rangan, and H. Zhang, {\it Green Communications: Theoretical Fundamentals, Algorithms, and Applications}, CRC Press, Sep. 2012.

\bibitem{HanAnsari2014}
T. Han and N. Ansari, ``Powering mobile networks with green energy,'' {\it IEEE Wireless Commun.}, vol. 21, no. 1, pp. 90-96, Feb. 2014.

{\color{black}
\bibitem{FRYu2012}
S. Bu, F. R. Yu, Y. Cai, and X. P. Liu, ``When the smart grid meets energy-efficient communications: green wireless cellular networks powered by the smart grid,'' {\it IEEE Trans. Wireless Commun.,} vol. 11, no. 8, pp. 3014-3024, Aug. 2012.
}

\bibitem{XueSmartGird}
X. Fang, S. Misra, G. Xue, and D. Yang, ``Smart grid - the new and improved power grid: a survey,'' {\it IEEE Commun. Surveys \& Tutorials}, vol. 14, no. 4, pp. 944-980, 2012.

\bibitem{SaadSPM}
W. Saad, Z. Han, H. V. Poor, and T. Basar, ``Game-theoretic methods for the smart grid: an overview of microgrid systems, demand-side management, and smart grid communications,'' {\it IEEE Sig. Process. Mag.}, vol. 29, no. 5, pp. 86-105, Sep. 2012.

\bibitem{Niu:CellZooming}
Z. Niu, Y. Wu, J. Gong, and Z. Yang, ``Cell zooming for
cost-efficient green cellular networks,'' {\it IEEE Commun. Mag.},
vol. 48, no. 11, pp. 74-78, Nov. 2010.

\bibitem{Goldsmith:Spectrum}
A. Goldsmith, S. A. Jafar, I. Maric, and S. Srinivasa, ``Breaking spectrum gridlock with cognitive radios: an information theoretic
perspective,'' {\it Proc. IEEE}, vol. 97, no. 5, pp. 894-914, May 2009.

\bibitem{Gesbert2010}
D. Gesbert, S. Hanly, H. Huang, S. Shamai, O. Simeone, and W. Yu, ``Multi-cell MIMO cooperative networks: a new look at interference,'' {\it IEEE J. Sel. Areas Commun.}, vol. 28, no. 9, pp. 1380-1408, Dec. 2010.

\bibitem{Gkatzikis2013}
L. Gkatzikis, I. Koutsopoulos, and T. Salonidis, ``The role of aggregators in smart grid demand response markets,'' {\it IEEE J. Sel. Areas Commun.}, vol. 31, no. 7, pp. 1247-1257, Jul. 2013.

\bibitem{XuGuoZhangGC2013}
J. Xu and R. Zhang, ``CoMP meets smart grid: a new communication and energy cooperation paradigm,'' to appear in {\it IEEE Trans. Veh. Tech.}. [Online]. Available: {\url{http://arxiv.org/abs/1303.2870}}.

\bibitem{GuoTCOM}
Y. Guo, J. Xu, L. Duan, and R. Zhang, ``Optimal energy and spectrum sharing for cooperative cellular systems,'' {\it IEEE Trans. Commun.}, vol. 62, no. 10, pp. 3678-3691, Oct. 2014.

\bibitem{XuZhangGC2014}
J. Xu and R. Zhang, ``Cooperative energy trading in CoMP systems powered by smart grids,'' to appear in {\it Proc. IEEE Globecom}, pp. 2738-2743, Dec. 2014.

\bibitem{Chia2013}
Y. K. Chia, S. Sun, and R. Zhang, ``Energy cooperation in cellular networks with renewable powered base stations,'' to appear in {\it IEEE Trans. Wireless Commun.}. [Online]. Available: {\url{http://arxiv.org/abs/1301.4786}}.



{\color{black}
\bibitem{TJLim2014}
J. Leithon, T. J. Lim, and S. Sun, ``Energy exchange among base stations in a cellular network through the smart grid,'' in {\it Proc. IEEE ICC}, pp. 4036-4041, Jun. 2014.
}
\end{thebibliography}
\end{document}